\documentstyle[psfig]{heliosymp}

\newcommand{\stt}{\small\tt}

\begin{opening}

\title{The Discovery of Non-Radial Gravity-mode Pulsations in $\gamma$ 
Doradus-Type Stars}

\author{K. KRISCIUNAS}
\institute{Department of Astronomy\\
University of Washington\\
Box 351580\\
Seattle, Washington 98195-1580 USA}

\end{opening}

\runningtitle{$\gamma$ Doradus-type Stars}

\begin{document}

{\fontsize{10pt}{11pt}\selectfont 

\begin{abstract}
Over two dozen early F-type variable stars have been identified 
which constitute a new class of pulsating stars.  These stars typically
have periods between 0.5 and 3 days with $V$-band variability of several
hundredths of a magnitude.  Given the time scale of the variability,
the pulsations would have to be non-radial gravity-mode pulsations.
The pulsational nature of some of these stars has been proven by means of
coordinated multi-longitude photometric campaigns, radial velocity (RV) 
variations and line profile (LP) variations, indicating low degree spherical
harmonics ($\ell$ = 3 or less).  Evidence is that these stars are younger
than 300 Myr; one would surmise that a rapid onset of convection in their
outer photospheric layers puts an end to the pulsations.  Further 
observation and modeling of these stars is important for our 
understanding of stellar evolution, the search for $g$-modes in the Sun, 
and is even relevant to the interpretation of radial velocity variations 
of solar-type stars (e.g. 51 Peg) in the search for extrasolar planets.

\end{abstract}

\section{Introduction}

Since the discovery by Cousins \& Warren (1963) that the early
F-type dwarf star $\gamma$ Doradus is variable, about two dozen
variable stars of similar spectral type and luminosity class have been 
identified which vary up to 0.1 mag on time scales much slower (e.g. 0.5 to
3 days) than the fundamental radial pulsational period (e.g. 1 to 3 
hours) for stars of this density .  Krisciunas \& Handler (1995) give a 
list of 17 {\em bona fide} $\gamma$ Dor stars and candidates.  A 
color-magnitude diagram of these and other stars is shown in Fig. 1.
Examples of light curves of two of the best studied examples are
found in Figs. 2 and 3. Updated information on these stars can be found 
at this website:

\vspace{3 mm}

{\stt http://www.ast.univie.ac.at/$^{\sim}$gerald/gdorlist.html}

\vspace{3 mm}

\begin{figure*}
\psfig{figure=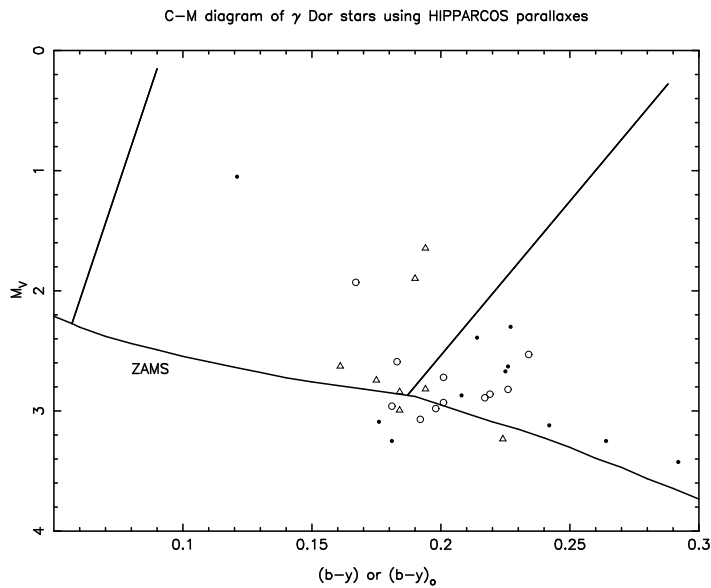,height=8cm,angle=-90}
\caption{Color-magnitude diagram of {\em bona fide} $\gamma$ Dor stars
(circles) and candidates (dots).  Stars in the open cluster NGC 2516 are
represented by triangles.  The position of the zero age main sequence and 
the borders of the $\delta$ Scuti instability strip are also indicated.
All but 3 of the field stars have absolute magnitudes derived from {\sc 
hipparcos} parallaxes.}
\end{figure*}

\begin{figure*}
\psfig{figure=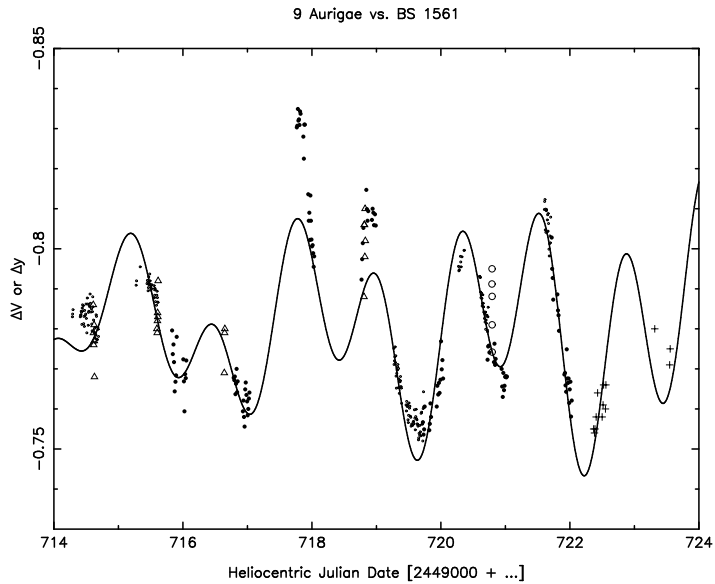,height=8cm,angle=-90}
\caption{Light curve of 9 Aurigae vs. BS 1561 during part of the 1994/5
campaign.  We fit the three frequencies found by Zerbi et al. (1997),
namely 0.7948, 0.7679, and 0.3429 d$^{-1}$.  Small dots represent data
by Garrido, Rodr\'{\i}guez, and Zerbi at Sierra Nevada Observatory, Spain.
Larger dots represent data by Krisciunas, Roberts, Crowe, and Pobocik at
Mauna Kea, Hawaii. Other data are by
Luedeke in Albuquerque, New Mexico (triangles), Guinan and McCook from Mt.
Hopkins, Arizona (open circles), and Sperauskas in Lithuania (+'s).
Note that even from day to day the amplitudes are irregularly variable.}  
\end{figure*}

\begin{figure*}
\psfig{figure=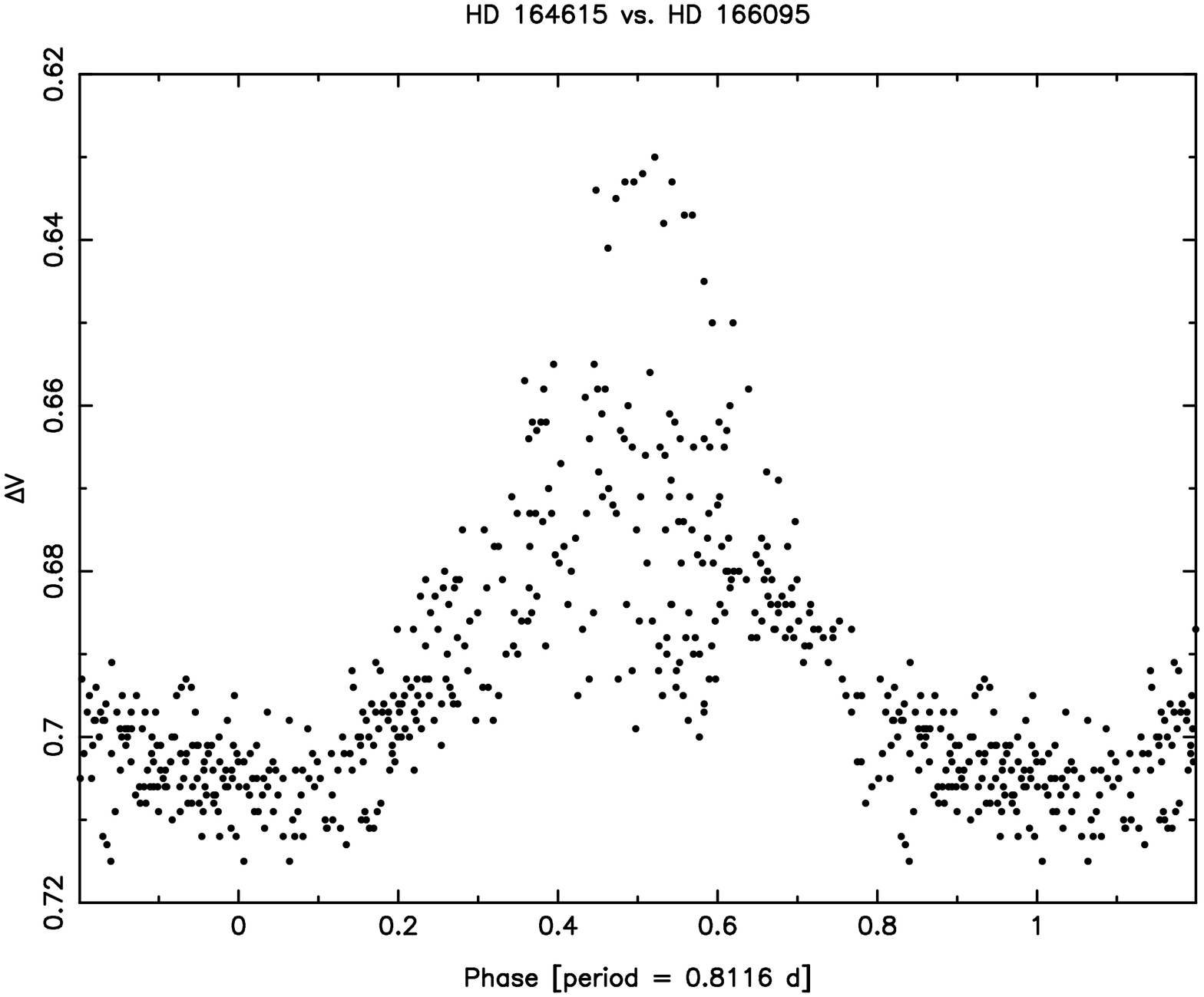,height=8cm,angle=-90}
\caption{Light curve of HD 164615 vs. HD 166095 during the 1995 campaign
(Zerbi et al. 1997).  These are the data from Sierra Nevada Observatory
only.  HD 164615 shows only one period, but the amplitude of the
photometric variations is clearly variable.}
\end{figure*}

As one can see in Fig. 1, the $\gamma$ Dor-type stars with {\sc 
hipparcos} parallaxes are found {\em on} the main sequence and overlap the
cool edge of the $\delta$ Scuti instability strip, an extension to
fainter absolute magnitudes of the classical Cepheid instability strip.
It would of course be of 
interest to delineate as accurately as possible the boundaries of this 
region of the HR diagram.

\section {What is causing the variability?}

Details of the light curve of a variable star (i.e. period, amplitude,
shape, color variations) usually provide sufficient clues to determine the
cause of the light variations.  Spectroscopic information (i.e. radial 
velocities $-$ RVs $-$ and line profiles $-$ LPs) can be used to prove one's 
case.  As Sherlock Holmes (1890) says: ``When you have eliminated the 
impossible, whatever remains, {\em however improbable}, must be the 
truth.''  This of course assumes that one can compile a {\em complete} 
list of suspects, or, in our case, make a complete list of causes of an 
observed phenomenon.

Eclipsing binaries show very regular, repeatable light curves, with one 
period and one or two minima per cycle (the primary one often being  
a magnitude deep).  $\gamma$ Dor stars typically
show multiple periods and a full range of a few hundredths to 0.1 mag in the 
Johnson $V$-band. We can easily state that $\gamma$ Dor stars are not 
eclipsing binaries.  

If we take the standard period-mean density equation for pulsating 
stars and express the fundmental radial pulsation period ($\Pi$) in 
terms of the radius and mass of the star in solar units, 
$\Pi \; = \; Q \; R_{\star}^{3/2} \; M_{\star}^{-1/2}$ .
$\gamma$ Dor stars are typically of spectral type F0 V and have 
$R_{\star} \approx 1.73 R_{\odot}$ (Poretti et al. 1997) and
$M_{\star} \approx 1.6  M_{\odot}$ (Popper 1980).  Q $\approx$ 0.033 days 
for the fundamental radial mode of $\delta$ Scuti stars (Fitch 1981, 
Table 2A), which are close cousins of the $\gamma$ Dor 
stars.  Thus, a $\gamma$ Dor star would have a fundamental radial
pulsation period of 1.4 hours, with a range of perhaps a factor of 2.
Radial overtones and non-radial pressure-mode pulsations would have 
periods shorter than this.

Given the time scales of photometric variability, $\gamma$ 
Dor stars are not pulsating in the fundamental 
radial mode, radial overtones, or by means of non-radial $p$-modes.  
While some are in binary or multiple systems, none is known to
have a close interacting companion.  One star {\em formerly} on the list 
has been shown to be an ellipsoidal primary of a binary system.  The northern
prototypical $\gamma$ Dor star, 9 Aurigae, was once thought to be a 
single-line spectroscopic binary with a period of 391.7 days, but that
interpretation of what are actually 2.89-day RV variations of
variable amplitude is definitely incorrect. 

After finding some ``slow'' variables with the colors of early F stars in
the cluster NGC 2516, Antonello \& Mantegazza (1986) suggested that 
early F-type stars could exhibit spots or non-radial gravity-mode 
oscillations.  Mantegazza {\em et al.} (1993) and Krisciunas (1994;
following discussions with Luis Balona and Jaymie Matthews) 
independently suggested that stars like $\gamma$ Dor constitute a new 
class of variables.  The former clearly favor rotational modulation of spots
(which could be dark or bright spots) as the explanation, while I advocated
that these were stars exhibiting non-radial $g$-modes.

Now, it is generally believed that stars with spectral types earlier than
F7 do not show evidence of star spots (Giampapa \& Rosner 1984). However,
G$\rm \ddot{u}$del, Schmitt \& Benz (1995) report the surprising result
that the F0V star 47 Cas exhibits evidence for strong coronal activity.
{\em If} a $\gamma$ Dor star were exhibiting rotational spot modulation,
three testable results can be checked: (1) Is the principal period
compatible with the size of the star and the projected equatorial
rotational speed?  (2) The RV minimum should be 90 degrees out of phase
with the luminosity maximum.  Finally, (3) spot models only work for stars with
one photometric period, or perhaps two {\em closely spaced} periods.

In the case of 9 Aur under the assumption of a spot model, the projected 
equatorial rotational speed and the principal photometric period imply 
that the star is within 16 degrees of being viewed pole-on (Krisciunas et
al. 1995a).  This means that there would be very little horizon beyond 
which any hypothetical spots could disappear.  Though one can invoke a spot
model with a variable projected area of spots as the star rotates, it would
be a rather contrived spot model that could account for variability up to
$\approx$ 0.1 mag with two or more {\em bona fide} periods.  (9 Aur has shown
evidence of 5 periods of variable amplitude.)

In the cases of 9 Aur and $\gamma$ Dor, which show a RV
range of about 4 km sec$^{-1}$, the most negative RV
more nearly {\em coincides} with the phase of minimum light of one 
of the demonstrable periods, rather than being 90 degrees out of phase as 
predicted by the spot model (Krisciunas et al. 1995a, Balona et al. 1996).

Aerts \& Krisciunas (1996) modeled 9 Aur as a non-radial gravity-mode
pulsator with $\ell$ = 3, $|m|$ = 1.  This was on the basis of 
{\sc coravel} autocorrelation diagrams obtained by Roger Griffin.  The
photometric variations are primarily driven at $f_1$ = 0.795 d$^{-1}$ while
the RV variations and LP variations are driven at the second most
significant photometric frequency of $f_2$ = 0.346 d$^{-1}$.
We found that, ``the amplitude of the radial part of the pulsation for $f_1$ 
is a factor of 4 larger than the one for $f_2$, while its angular 
dependence is the same. Since the photometric variability is determined 
most of all by the radial part of the pulsation, it is quite 
understandable that the photometric variability is dominated by the mode 
with frequency $f_1$.''

Balona et al. (1994) tried to model $\gamma$ Dor itself by means of a
differentially rotating spotted star model, but concluded that non-radial
$g$-modes were a better explanation.  Since the confirmation of a third
frequency for this star and the analysis of LP and RV
variations (Balona et al. 1996), the spot model is rejected
and the $g$-mode model is confirmed.

A third $\gamma$ Dor star which has been confirmed to be a
non-radial $g$-mode pulsator is HD 164615 (Zerbi et al. 1997; Hatzes, 
in preparation).  It exhibits photometric variations most likely
at only a single frequency of 1.2321 d$^{-1}$, a frequency
which has been stable for over 10 years.  Like other $\gamma$ Dor stars, 
it has a variable photometric amplitude.  Its LP variations 
(see Fig. 4) can be modeled by an $\ell$ = 2 sectoral mode non-radial 
pulsator with a mean pulsation amplitude of $\approx$ 7 km sec$^{-1}$.  
The pulsation amplitude seems to be a function of phase.  $\ell$ = 3 and 
4 can also fit the profiles, but the amplitude of the pulsations is lower.

Mantegazza et al. (1994) provide some evidence that the 
multi-periodic $\gamma$ Dor star HD 224638 also shows LP 
variations.

\begin{figure*}
\psfig{figure=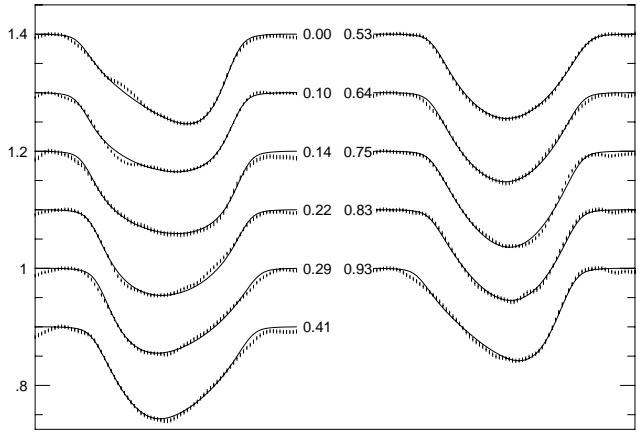,height=8cm,angle=90}
\caption{Profiles of the Fe II 531.8 nm line in HD 164615, measured by
Hatzes.  These are single spectra with phases based on an  epoch of
JD 2449521.710 and a period of 0.8116 d (Zerbi et al. 1997).  The 
solid lines are fits based on an $\ell$ = 2 sectoral mode model with 
pulsation amplitude 7 km sec$^{-1}$.}
\end{figure*}

\section {The $\gamma$ Dor phenomenon is related to age}

Eggen (1995) and Krisciunas et al. (1995b; following some discussion with
Balona) independently suggested that $\gamma$ Dor stars are all 
relatively young.  Many of the field stars have 
space velocities like young disk stars.  $\gamma$ Dor is embedded in
a $\beta$ Pictoris-like disk or envelope.  Eight candidates are found in the
cluster NGC 2516 (age 120 $\pm$ 20 Myr), one in the Pleiades (age $\approx$
80 Myr), and one in M 34 (Krisciunas \& Crowe 1997), whose age is $\approx$
250 Myr.  Krisciunas et al. (1995b) found no $\gamma$ Dor stars in the
Hyades,
whose age from {\sc hipparcos} data is 625 $\pm$ 50 Myr (Perryman et al.
1997).  Eggen suggests that the $\gamma$ Dor stars all lie in the
``$\rm B\ddot{o}hm$-Vitense decrement'', a gap in the main sequence of many
star clusters at T$_{eff} \approx$ 7700 K.  $\gamma$ Dor stars have
such temperatures and presumably have not yet experienced a rapid onset of 
convection in their photospheres.  Once convection sets in, the 
pulsations presumably cease.  Modeling of $\gamma$ Dor stars, however,
has not yet been accomplished to any satisfactory degree.   Gautschy \& $\rm 
L\ddot{o}ffler$ (1996) were unable to find overstable low-degree
oscillation modes for purely radially symmetric stars without any 
interaction with convection, rotation, or magnetic fields.  

If we are correct that $\gamma$ Dor stars are all younger than $\approx$
300 Myr $-$ their main sequence lifetimes are $\approx$ 3 Gyr $-$ then we 
now have a means of obtaining an upper bound to the ages of some 
stars which are not in clusters.  Also, this provides a means of determining
an upper limit to the ages of some white dwarfs, such as the E-component 
of 9 Aur (Krisciunas et al. 1993), a star whose companionship to its 
supposed primary should be confirmed.

\section {Discussion}

Many types of variable stars are characterized by specific types of
observational evidence (i.e. light curve amplitude and shape, period(s),
correlations of photometric color and brightness, RV
variations, LP variations, and signatures at non-optical 
wavelengths such as infrared or ultraviolet excesses).  We of course want
to understand the {\em physical} mechanism for a star's behavior.  

The sum total of observational evidence on $\gamma$ Dor stars indicates
that the most likely explanation of their behavior is non-radial
gravity-mode oscillations.  This conclusion is based on their
typically multiple periods, the time scale and amplitude of such 
variations, and the correlation (including phase) of those variations 
with RV and LP variations.  Early F-stars with
one period (even if of variable amplitude) could still be modeled by
means of a (bright) spot model.  However, it would be a cruel trick of 
Nature, in my opinion, if some stars in the $\gamma$ Dor region of the
HR Diagram were variable due to $g$-modes, while others were variable 
due to spots.
  
One of the most exciting discoveries of the past two years has been
evidence for extrasolar planets (Boss 1996).  All we know {\em for certain}
is that there are solar-type stars with  RV variations like
those expected {\em if} there are Jupiter-mass planets orbiting them.
Some of these variations are on time scales comparable to that
of $\gamma$ Dor stars, and there is {\em marginal} evidence that 51 Peg 
exhibits LP variations $-$ something that should {\em not} be 
observed if the RV variations of the star are due to an 
orbiting planet (Gray 1997; but see also Hatzes et al. 1997).  

Consider for a moment the RV variations of 9 Aurigae, which in
1993/4 followed its 2.9 d period and which had an amplitude of 2.00 km
sec$^{-1}$.  The corresponding photometric amplitude was 12.2 mmag.  
51 Peg has a RV amplitude of 59 m sec$^{-1}$ (Mayor \& Queloz 1995).  
{\em If} 51 Peg were exhibiting
non-radial pulsations like a scaled down version of 9 Aurigae,
the corresponding photometric amplitude would be less than 4 {\em
ten}-thousandths of a magnitude, clearly beyond our detection
capabilities, the claims of Henry et al. (1997) notwithstanding.  Thus, if
we were to demonstrate that a 51 Peg-type star were pulsating, it would
have to be on the basis of RV and LP variations.

Now the question arises, and Tim Bedding brought it up after one of
the talks at this Symposium: Can the presence of a close planet
tidally induce non-radial $g$-modes in solar-type stars? Terquem
et al. (1998) have looked into this question and found that a companion
orbiting 51 Peg with a period of 4.23 d induces a radial velocity at
the stellar surface, the maximum of which is between 10$^{-2}$ and
6 m sec$^{-1}$ for a companion mass between 10$^{-3}$ and 1.0 M$_{\odot}$.
Such minuscule induced RV variations are not observable for Jupiter-like
planets.  Because
of the extreme predictability of the RV variations in 51 Peg-type
stars and the ragged RV variations for non-radial pulsators, the planetary
would hypothesis would be the more likely of the two for 51 Peg-type
stars.  However, because Nature is more devious than we are,
we must obtain high resolution spectra of 51 Peg-type stars to see
if any do show LP variations.  If they do, such variations could be due to
the long-sought $g$-modes in stars as cool as the Sun.

\end{document}